\documentclass[hyper]{JHEP} 

\usepackage{epsfig}

\newcommand\fverb{\setbox\pippobox=\hbox\bgroup\verb}

\newcommand\fverbdo{\egroup\medskip\noindent%

            \fbox{\unhbox\pippobox}\ }

\newcommand\fverbit{\egroup\item[\fbox{\unhbox\pippobox}]}

\newbox\pippobox

\title{Note About String with Deformed Dispersion Relation}
\author{J. Kluso\v{n}\\
Department of
Theoretical Physics and Astrophysics\\
Faculty of Science, Masaryk University\\
Kotl\'{a}\v{r}sk\'{a} 2, 611 37, Brno\\
Czech Republic\\
E-mail: \email{klu@physics.muni.cz}}
\preprint{}

 \abstract{We study string theory with 
global momentum living on de Sitter space. We also show that this presumption leads to the string with  deformed dispersion 
 relation.}

\def\be{\begin{equation}}

\def\tp{\tilde{p}}
\def\tx{\tilde{x}}

\def\ee{\end{equation}}

\def\bea{\begin{eqnarray}}

\def\eea{\end{eqnarray}}

\def\bJ{\mathbf{J}}

\def\tmH{\tilde{\mH}}

\def\mH{\mathcal{H}}

\def\mC{\mathcal{C}}

\newcommand{\mG}{\mathcal{G}}

\newcommand{\bT}{\mathbf{T}}

\newcommand{\mL}{\mathcal{L}}

\def\tJ{\tilde{J}}

\def\bS{\mathbf{S}}
\def\pb #1{\left\{#1\right\}}

\begin{document}
\section{Introduction and Summary}\label{first}
There was an intensive study of the particle in the curved momentum space going back to 60's when these theories were studied as possible examples of divergence free theories
\cite{Golfand:1959vqx}.  Another very interesting situation was analysed in 
\cite{Arzano:2010kz} and in 
\cite{Girelli:2005dc} where the particle with the four dimensional de Sitter space with curvature $\kappa$ was investigated. Explicitly, in term of five dimensional Minkowski space-time coordinates $P_A$ it has the form 
\begin{equation}
-P_0^2+P_1^2+P_2^2+P_3^2+P_4^2=\kappa^2 \ . 
\end{equation}
Further  the Casimir for the first four coordinates forming $SO(3,1)$ subgroup of $SO(4,1)$ that leaves the quadratic form given above invariant is 
\begin{equation}
\mC=P_0^2-P_1^2-P_2^2-P_3^2-m^2 \ . 
\end{equation}
With these two main ingrediences the authors in \cite{Arzano:2010kz} derived an action 
for massive particle with de Sitter momentum space. It was shown that this action has
deformed dispersion relation and that it possesses non-trivial symplectic structure. 

We would like to see whether similar idea can be applied in case of string theory where the fundamental object is two dimensional string instead of one dimensional point particle. Explicitly, we consider $D-$dimensional string theory and impose the condition that momenta lives on $D-$dimensional de Sitter space. With this presumption we expect
that we get string theory with deformed disperssion relation whose existence was firstly proposed in \cite{Magueijo:2004vv}. We studied  this proposal from Hamiltonian point of view in \cite{Kluson:2021whs} and we showed that the formulation of string theory 
presented in \cite{Magueijo:2004vv} cannot lead to string theory with deformed dispersion relation. We also suggested in \cite{Kluson:2021whs} that the way how to define string 
with deformed dispersion relation is to follow analysis performed in 
\cite{Girelli:2005dc,Arzano:2010kz}. As we argued above
we have to impose the condition that momenta lives on $D-$ dimensional de Sitter space. However the question is whether this constraint should restrict momentum density that depends on world-sheet spatial coordinate. We argue that such constraint cannot be imposed due to the fact that its Poisson bracket with spatial diffeomorphism constraint does not vanish on the constraint surface that would imply that they are not the first class constraints breaking consistency of the string theory. For that reason we should rather impose the condition that the global momentum is restricted to live on 
$D-$ dimensional de Sitter space. In other words we consider string with additional first class global constraint. Then, following \cite{Arzano:2010kz}, we fix this global constraint and introduce appropriate variables that parametrize reduced phase space. As a consequence of this procedure we find string theory with modified dispersion relation between global momenta. Note that since the inner part of the string is not affected by
this modification clearly consistency and the spectrum of this string is the same as in case of undeformed one. In other words we can very easily implement the modification of dispersion relation into string theory. Then we can study properties of such a modified
string theory in the same way as in point particle with all consequences. 

This paper is organized as follows. In the next section (\ref{second}) we introduce
string theory with global momentum to be restricted on de Sitter manifold. Then in section (\ref{third}) we study its canonical form and determine corresponding Dirac brackets.

\section{String Theory with Deformed Dispersion Relation}\label{second}
Our starting point is the action for bosonic  string in $D-$dimensions. 
In order to find its form with deformed dispersion relation it is convenient
to write it in the canonical form together with an additional constraint whose
explicit form will be specified below. Explicitly, we consider  string theory action  in the  form
\begin{equation}
S=\int d\tau d\sigma (p_A\partial_\tau x^A-N^\tau \mH_\tau-
N^\sigma \mH_\sigma -\Gamma \mathcal{S}) \ , 
\end{equation}
where $\mH_\tau\approx 0 \ , \mH_\sigma\approx 0$ are standard Hamiltonian
and diffeomorphism constraints of bosonic string in the form 
\begin{equation}
\mH_\tau=p_\mu \eta^{\mu\nu}p_\nu+T^2 \partial_\sigma x^\mu \eta_{\mu\nu}\partial_\sigma x^\nu \quad \mH_\sigma=p_\mu\partial_\sigma x^\mu \ , 
\end{equation}
where $\mu,\nu=0,1,\dots,D$ and where $A,B=0,1,\dots,D+1$. We further presume
closed  string with the length $L$.
%
%
An important ingredient is the constraint that enforces momentum to be de Sitter
\begin{equation}
\mathcal{S}=p_A p^A-\kappa^2=p_\mu p^\mu+p^2_{D+1} -\kappa^2\approx 0 \ . 
\end{equation}
This is natural generalization of the constraint used in \cite{Arzano:2010kz}. However 
an important point is that this  local constraint $\mathcal{S}$ does not have weakly vanishing Poisson bracket with $\mH_\sigma \approx 0$
on the constraint surface. Such a situation would imply that $\mH_\tau,\mH_\sigma$ were not the first class constraints with the breaking of the consistency of string theory. 
For that reason we would rather presume that the total momentum is forced to live on de Sitter space and that $p_{D+1}$ depends on $\tau$ only. Explicitly we consider following constraint
\begin{equation}
\bS=(\int d\sigma p_\mu)(\int d\sigma p^\mu)+p_{D+1}^2-\kappa^2\approx 0 \ . 
\end{equation}
In order to demonstrate consistency of the theory let us introduce  smeared form of spatial diffeomorphism and Hamiltonian constraints
\begin{equation}
\bT_\sigma(N^\sigma)=\int d\sigma N^\sigma \mH_\sigma \ , \quad 
\bT_\tau(N)=\int d\sigma N \mH_\tau \ . 
\end{equation}
Then using canonical  Poisson brackets $\pb{x^\mu(\sigma),p_\nu(\sigma')}=
\delta^\mu_\nu\delta(\sigma-\sigma')$
we obtain following Poisson brackets between all constraints
\begin{eqnarray}\label{PBcons}
& &\pb{\bT_\sigma (N^\sigma),\bS}=0 \ , \quad 
\pb{\bT_\tau(N),\bS}=0 \ , \nonumber \\
& &\pb{\bT_\sigma(N^\sigma),\bT_\sigma (M^\sigma)}=\bT_\sigma(N^\sigma \partial_\sigma M^\sigma-M^\sigma \partial_\sigma N^\sigma) \ , \nonumber \\
& &\pb{\bT_\sigma(N^\sigma),\bT_\tau(M)}=
\bT_\tau(-\partial_\sigma N^\sigma M+N^\sigma \partial_\sigma M) \ , \nonumber \\
& &\pb{\bT_\tau(N),\bT_\tau(M)}=4\bT_\sigma(N\partial_\sigma M-M\partial_\sigma N) \ . 
\nonumber \\
\end{eqnarray}
 The formulas given in (\ref{PBcons}) show
that all Poisson brackets between constraints vanish on the constraint surface and
hence they are the first class constraints. Hence the action has the form 
\begin{equation}
S=\int d\tau d\sigma (p_A
\partial_\tau x^A-N^\tau \mH_\tau-N^\sigma \mH_\sigma)-\int
d\tau \Gamma \bS \ . 
\end{equation}
Since $\bS$ is the first class constraint it is appropriate to fix it. In fact, since
$\bS\approx 0 $  is global constraint that does not depend on $\sigma$ we can presume that its gauge fixing eliminates global variables only. For that reason we split $p_\mu$ and $x^\mu$ in the following way
\begin{equation}\label{pmusplit}
p_\mu(\tau,\sigma)=\tp_\mu(\tau,\sigma)+\frac{1}{L}P_\mu(\tau) \ , \quad 
x^\mu(\tau,\sigma)=\tx^\mu(\tau,\sigma)+X^\mu (\tau) \ , 
\end{equation}
where
\begin{equation}
P_\mu=\int d\sigma p_\mu \ , \quad X^\mu=\frac{1}{L}\int d\sigma x^\mu \ ,
\end{equation}
where $L$ is the length of the string and we have chosen the pre factor
$L^{-1}$ in the expression for $X^\mu$ so that it has correct dimension $[length]$.
Note that (\ref{pmusplit}) implies 
\begin{equation}
\int d\sigma \tp_\mu=0 \ , \quad \int d\sigma \tx^\mu=0 \ .
\end{equation}
Further, using (\ref{pmusplit}) and the canonical Poisson brackets $\pb{x^\mu(\sigma),
p_\nu(\sigma')}=\delta^\mu_\nu\delta(\sigma-\sigma')$ 
we obtain
\begin{eqnarray}
& &\pb{X^\mu,P_\nu}=\delta^\mu_\nu \ , \quad \pb{\tx^\mu(\sigma),P_\nu}=0 \ , \nonumber \\ 
& &\pb{X^\mu,\tp_\nu(\sigma)}=0 \ , \quad  \pb{\tx^\mu(\sigma),\tp_\nu(\sigma')}=
\delta(\sigma-\sigma')\delta^\mu_\nu+\frac{1}{L}\delta^\mu_\nu \ .  \nonumber \\
\end{eqnarray}
Finally, using (\ref{pmusplit}) we can rewrite the Hamiltonian constraint and spatial diffeomorphism constraint in the form 
\begin{eqnarray}
& &\mH_\tau=\tp_\mu \eta^{\mu\nu}\tp_\nu+T^2\partial_\sigma \tx^\mu\eta_{\mu\nu}
\partial_\sigma \tx^\nu+\frac{1}{L}P_\mu \eta^{\mu\nu}P_\nu 
\equiv\frac{1}{L}P_\mu \eta^{\mu\nu}P_\nu+\tmH_\tau  \  , \nonumber \\
& &\mH_\sigma=\tp_\mu\partial_\sigma \tx^\mu +\frac{1}{L}P_\mu\partial_\sigma \tx^\mu\equiv 
\tmH_\sigma+\frac{1}{L}P_\mu\partial_\sigma \tx^\mu \ . 
\nonumber \\
\end{eqnarray}
As we argued above $\bS$  Poisson commutes with all constraints so it is appropriate
to gauge fix it. It is important to stress that the innter variables $\tx^\mu,\tp_\mu$ are not affected by this procedure.  Following \cite{Arzano:2010kz} we introduce gauge fixing
function $\mG$ that does not depend on $\sigma$ and that obeys
\begin{equation}
\pb{\bS,\mG}\equiv \triangle\neq 0 \ 
\end{equation}
so that $\bS$ and  $\mG$ are two second class constraints and hence we reduce
global degrees of freedom from $2(D+1)$ into $2D$. Following \cite{Arzano:2010kz} we parametrize
reduced phase space with variables $y^\mu,k_\mu$ that have vanishing Poisson brackets with constraints
\begin{eqnarray}
\pb{k_\mu,\bS}\approx 0 \ , \quad  \pb{k_\mu,\mG}\approx 0 \ , \nonumber \\
\pb{y^\mu,\bS}\approx 0\ ,  \quad \pb{y^\mu,\mG}\approx 0 \ . 
\nonumber \\
\end{eqnarray}
Now we find that Dirac brackets of $k_\mu$ and $y^\mu$ with arbitrary functions
are equal to Poisson brackets since
\begin{eqnarray}
\pb{y^\mu,F}_D=\pb{y^\mu,F}-\pb{y^\mu,\bS}\triangle^{-1}\pb{\mG,F}+
\pb{y^\mu,\mG}\triangle^{-1}\pb{\bS,F}=\pb{y^\mu,F} \ , 
\nonumber \\
\pb{k_\mu,F}_D=\pb{k_\mu,F}-\pb{k_\mu,\bS}\triangle^{-1}\pb{\mG,F}+
\pb{k_\mu,\mG}\triangle^{-1}\pb{\bS,F}=\pb{k_\mu,F} \ .
\nonumber \\
\end{eqnarray}
As the next step we introduce gauge fixing function $\mG$ 
that, following \cite{Arzano:2010kz}, has the form
\begin{equation}
\mG=\frac{1}{2}(X^{D+1}-X^0)-T \ , 
\end{equation}
where $T$ is real number. Then we have
\begin{equation}
\pb{\bS,\mG}=-(P_0+P_{D+1}) \ . 
\end{equation}
In what follows we will consider $AN(D)-$ group submanifold defined by two conditions $\bS=0 \ , P_0+P_{D+1}>0$ . Let us choose $k_\mu$ coordinates in the following way
\begin{eqnarray}\label{defP}
& &P_0(k_0,k_i)=\kappa \sinh\frac{k_0}{\kappa}+\frac{k_i k^i}{2\kappa}e^{\frac{k_0}{\kappa}} \ , \nonumber \\
& &P_i(k_0,k_i)=k_ie^{\frac{k_0}{\kappa}} \ , \nonumber \\
& &P_{D+1}(k_0,k_i)=\kappa \cosh \frac{k_0}{\kappa}-\frac{k_i k^i}{2\kappa}e^{\frac{k_0}{\kappa}} \ , 
\nonumber \\
\end{eqnarray}
where $i=1,\dots,D$. Then using the first and the third equation in 
(\ref{defP}) we can express $k_0$ as
\begin{equation}
k_0=\kappa\ln\frac{P_0+P_{D+1}}{\kappa} \ . 
\end{equation}
Inserting this relation into the second equation in (\ref{defP}) we can express
$k_i$ as 
\begin{equation}
k_i=
\kappa \frac{P_i}{P_0+P_{D+1}} \ . \end{equation}
In other words we have $k_\mu=k_\mu(P_A)$. Then using these relations we immediately
get 
\begin{equation}
\pb{k_\mu,\bS}=0  \ , \quad 
\pb{k_0,\mG}=0 \ .
\end{equation}
In the same way we can show that $\pb{k_i,\mG}=0$ since $k_i$ depend on 
the linear combination $P_0+P_{D+1}$ only. We can further argue that 
\begin{equation}
\pb{k_\mu,\tx^\nu}=0
\end{equation}
due to the fact that $k_\mu$ are functions of $P_\mu$ and $P_\mu$ Poisson commute
with $\tx^\mu$ by definition. 
To proceed further we introduce generator of rotations that for the full string has the form
\begin{eqnarray}
& &J_{AB}=\int d\sigma (x_Ap_B-x_Bp_A)=\nonumber \\
& &(X_AP_B-X_BP_A)+\int d\sigma (\tx_A\tp_B-\tx_B\tp_A)=
\bJ_{AB}+\tJ_{AB} \ .  \nonumber \\
\end{eqnarray}
Then it is easy to see that 
\begin{equation}
\pb{J_{AB},\bS}=0 
\  .
\end{equation}
Following \cite{Arzano:2010kz} we define $y^\mu$ using components of $J_{AB}$  as
\begin{equation}\label{defy}
y^0=\frac{1}{\kappa}\bJ_{0(D+1)} \ , \quad 
y^i=\frac{1}{\kappa}(\bJ_{(D+1)i}+\bJ_{(D+1)i})
\end{equation}
and calculate (recall that $X_0=-X^0,X_{D+1}=X^{D+1}$)
\begin{equation}
\pb{y^0,\mG}=\frac{1}{2}(X^0-X^{D+1})=- (\mG+T)\approx -T
\end{equation}
and we see that this Poisson bracket is zero for $T=0$. Let us further calculate
Poisson bracket between $y^i$ and $\mG$ and we get
\begin{eqnarray}
\pb{y_i,\mG}=0 \ , \nonumber \\
\end{eqnarray}
where we used  the fact that we have
\begin{eqnarray}\label{pbbJAB}
\pb{\bJ_{AB},\bJ_{CD}}=
(\eta_{AD} \bJ_{CB}+\eta_{BC}\bJ_{DA}+\eta_{AC}\bJ_{BD}+\eta_{BD}
\bJ_{AC}) \ . 
\nonumber \\
\end{eqnarray}
Then with the help of definition (\ref{defy}) and the Poisson bracket
(\ref{pbbJAB}) we get
\begin{eqnarray}
& &\pb{y^i,y^j}=\frac{1}{\kappa^2}(\pb{\bJ_{0i},\bJ_{0j}}+\pb{\bJ_{0i},\bJ_{(D+1)j}}+
\pb{\bJ_{(D+1)i},\bJ_{0j}}+\pb{\bJ_{(D+1)i},\bJ_{(D+1)j}})=0 \ , \nonumber \\
& &\pb{y^0,y^i}=\frac{1}{\kappa^2}\pb{\bJ_{0(D+1)},\bJ_{0i}+\bJ_{(D+1)i}}=-\frac{1}{\kappa^2} 
(\bJ_{0i}+\bJ_{(D+1)i})=- \frac{1}{\kappa} y^i \ . \nonumber \\
\end{eqnarray}
If we finally take into account the gauge fixing function $\mG=0$ we can find
inverse relation between $X^A$ and $y^\mu$ in the form
\begin{eqnarray}
& &X^{D+1}=-\frac{\kappa y^0}{P_{D+1}+P_0}=-y^0 e^{-\frac{k_0}{\kappa}} \ , \nonumber \\ & &X^0=-\frac{\kappa y^0}{P_{D+1}+P_0}=-y^0e^{-\frac{k_0}{\kappa}} \ ,  \nonumber \\ 
& &X^i=-\frac{\kappa y^i}{P_0+P_{D+1}}=-x^ie^{-\frac{k_0}{\kappa}} \ . \nonumber \\ 
\end{eqnarray}
Then inserting these relations into string theory action we obtain
\begin{eqnarray}\label{Sfinaldef}
S=\int d\tau (\dot{y}^\mu k_\mu-\frac{\dot{k}_0}{\kappa}y^i k_i)
+\int d\tau d\sigma (\tp_\mu\partial_\tau \tx^\mu-
N(P_\mu (k)P^\mu(k)+\tmH_\tau)-N^\sigma \mH_\sigma(k)) \ ,
\nonumber \\
\end{eqnarray}
where
\begin{equation}
P_\mu(k)P^\mu(k)=-4\kappa^2\sinh^2\left(\frac{k_0}{2\kappa}\right)+k_i k^ie^{\frac{k_0}{\kappa}} \ .
\end{equation}
The action (\ref{Sfinaldef}) is final form of the string theory action with modified dispersion relation. We see that the inner part has the same form as in non-deformed case. In the next section we perform canonical analysis of the action (\ref{Sfinaldef}). 
\section{Hamiltonian Analysis}\label{third}
In this section we perform canonical analysis of the action (\ref{Sfinaldef}), following 
\cite{Arzano:2010kz}. To do 
this we  tread $y^\mu$ and $k_\mu$ as independent variables. In other words we can interpret
the action (\ref{Sfinaldef}) as Lagrangian form of the action and determine conjugate
momenta from it. 
 We define $l_\mu$ as momentum conjugate to $y^\mu$ an $\Pi^\mu
$ as momentum conjugate to $k_\mu$
\begin{eqnarray}\label{defmom}
l_\mu=\frac{\partial \mL}{\partial \partial_\tau y^\mu}=k_\mu \ , \quad 
\Pi^i=\frac{\partial \mL}{\partial \partial_\tau k_i}=0 \ , 
\quad 
\Pi^0=\frac{\partial \mL}{\partial \partial_\tau k_0}=-\frac{1}{\kappa}y^ik_i \ , 
\nonumber \\
\end{eqnarray}
where we have  canonical Poisson brackets
\begin{equation}\label{canPB}
\pb{y^\mu,l_\nu}=\delta^\mu_\nu \ , \quad
\pb{k_\mu,\Pi^\nu}=\delta_\mu^\nu \ . 
\end{equation}
Further, from  (\ref{defmom}) we get following primary 
constraints
\begin{eqnarray}
\phi_\mu\equiv l_\mu-k_\mu\approx 0 \ , \quad 
\Psi^i\equiv \Pi^i\approx 0 \ , \quad \Psi^0=\Pi^0+
\frac{1}{\kappa} y^i k_i \ . \nonumber \\
\end{eqnarray}
Then with the help of (\ref{canPB}) we obtain  Poisson brackets between constraints in the form 
\begin{eqnarray}\label{pbcons}
& &\pb{\phi_0,
\Psi^0}=-1 \ , \quad  \pb{\phi_0,\Psi^i}=0 \ , \quad 
\pb{\phi_i,\Psi^0}=-\frac{1}{\kappa}k_i \ , \nonumber \\
& &\pb{\phi_i,\Psi^j}=-\delta_i^j \ , \quad 
\pb{\Psi^i,\Psi^0}=-\frac{1}{\kappa}y^i \ . \nonumber \\
\end{eqnarray}
It is convenient to write these Poisson brackets into matrix form introducing common
notation for all constraints $\Phi_A\equiv(\phi_0,\phi_i,\Psi^0,\Psi^i)$ and hence
we can rewrite (\ref{pbcons}) into more symmetric form
\begin{equation}\label{MAB}
\pb{\Phi_A,\Phi_B}=M_{AB} \ , \quad 
M_{AB}=\left(\begin{array}{cccc} 
0 & 0 &  -1 & 0 \\
0 & 0 & -\frac{1}{\kappa}k_i & -\delta_i^j \\
1 & \frac{1}{\kappa}k_j & 0 & \frac{1}{\kappa}y^j
\\
0 & \delta_{i}^j & -\frac{1}{\kappa}y^i & 0 \\
\end{array}\right)  \ .  
\end{equation}
Then it is easy  to determine inverse matrix $M^{AB}$ from (\ref{MAB}) 
\begin{equation}
M^{AB}=\left(\begin{array}{cccc} \\
0 & \frac{1}{\kappa}y^j & 1 & \frac{1}{\kappa}k_j \\
-\frac{1}{\kappa}y^i & 0 & 0 & \delta_i^j \\
-1 & 0 & 0 & 0 \\
-\frac{1}{\kappa}k_i & -\delta_i^j & 0 & 0 \\
\end{array}\right) \ . 
\end{equation}
Now we are ready to determine  Dirac brackets between canonical variables 
(Note that due to the fact that $\Phi_A$ are second class constraints
we can solve them for $\Pi^\mu$ and $l_\mu$ so that independent variables
will be $y^\mu$ and $k_\mu$ ). First of all we have
\begin{eqnarray}
& &\pb{y^0,\Phi_A}=(1,0,0,0) \ , \quad 
\pb{y^i,\Phi_A}=(0,\delta^i_j,0,0) \ , \nonumber \\ 
& &\pb{k_0,\Phi_A}=(0,0,1,0) \ , \quad  \pb{k_i,\Phi_A}=(0,0,0,\delta_i^j)
\nonumber \\
\end{eqnarray}
so that we have
\begin{eqnarray}
& &\pb{y^i,y^j}_D=\pb{y^i,y^j}-\pb{y^i,\Phi_A}M^{AB}\pb{\Phi_B,y^j}=0 \ , 
\nonumber \\
& &\pb{y^0,y^i}_D=\pb{y^0,y^i}-\pb{y^0,\Phi_A}M^{AB}\pb{\Phi_B,y^i}=
\frac{1}{\kappa}y^i \ , \nonumber \\
& &\pb{y^0,k_0}_D=\pb{y^0,k_0}-\pb{y^0,\Phi_A}M^{AB}\pb{\Phi_B,k_0}=1 \ , 
\nonumber \\
& &\pb{y^0,k_i}_D=\pb{y^0,k_i}-\pb{y^0,\Phi_A}M^{AB}\pb{\Phi_B,k_i}=
\frac{1}{\kappa}k_i \ , \nonumber \\
& &\pb{y^i,k_0}_D=\pb{y^i,k_0}-\pb{y^i,\Phi_A}M^{AB}\pb{\Phi_B,k_0}=0 \ , 
\nonumber \\
& &\pb{y^i,k_j}_D=\pb{y^i,k_j}-\pb{y^i,\Phi_A}M^{AB}\pb{\Phi_B,k_j}=\delta^i_j \ , 
\nonumber \\
& &\pb{k_0,k_i}_D=\pb{k_0,k_i}-\pb{k_0,\Phi_A}M^{AB}\pb{\Phi_B,k_i}=0 \ , \nonumber \\
& &\pb{k_i,k_j}_D=\pb{k_i,k_j}-\pb{k_i,\Phi_A}M^{AB}\pb{\Phi_B,k_j}=0 \ . \nonumber \\
\end{eqnarray}
These Dirac brackets are the same as in the point particle case derived in 
\cite{Arzano:2010kz} and that are counterparts of $\kappa-$deformed phase space
\cite{Lukierski:1993wx}. Once again we should stress  that $k_\mu,y^\mu$ have zero Dirac brackets with $\tx^\mu,\tp_\mu$. In fact, there are still Hamiltonian and diffeomorphism constraints
\begin{eqnarray}
\mH_\tau=P_\mu(k)\eta^{\mu\nu}P_\nu(k)+\tmH_\tau \ , \mH_\sigma=\tmH_\sigma+P_\mu(k)\partial_\sigma \tx^\mu
\nonumber \\
\end{eqnarray}
that are still first class constraints. Finally, since inner part of the string 
represented by variables $\tx^\mu,\tp_\mu$ decouple from $k_\mu,y^\mu$ it is clear
that the quantum consistency of the string is the same as in undeformed case. 

Let us outline main result that was derived in this paper. The goal was to find
string theory with deformed dispersion relation in an alternative way to the proposal
presented in \cite{Magueijo:2004vv}. We showed that it can be done when we generalized
procedure suggested in \cite{Arzano:2010kz,Girelli:2005dc} from one dimensional object
to two dimensional object-string.

{\bf Acknowledgement:}
\\
This work 
is supported by the grant “Integrable Deformations”
(GA20-04800S) from the Czech Science Foundation
(GACR).

\end{document}